\documentclass[a4paper,11pt]{article} 
\pdfoutput=1 
\usepackage{lineno}
\usepackage{comment}
\usepackage{graphicx}
\usepackage{xcolor}
\usepackage[affil-it]{authblk} 
\usepackage{amsmath}

\title{Calibration of Solid State Nuclear Track Detectors for Rare Event Searches}

\author[1]{M.~Kalliokoski}
\author[2]{G.~Levi}
\author[2,3]{A.~Maulik\thanks{Corresponding author: atanu.maulik@bo.infn.it}}
\author[4]{I.~Ostrovskiy}
\author[2]{L.~Patrizii}
\author[3]{J.~Pinfold}
\author[2]{Z.~Sahnoun\thanks{Corresponding author: sahnoun@bo.infn.it}}
\author[2]{G.~Sirri}
\author[3]{R.~Soluk}
\author[5]{M.~Staelens}
\author[2]{V.~Togo}
\author[4]{A.~Upreti}

\affil[1]{\small Helsinki Institute of Physics, University of Helsinki, Helsinki, Finland}
\affil[2]{INFN Section of Bologna and DIFA University of Bologna, Bologna, Italy}
\affil[3]{Physics Department, University of Alberta, Edmonton, Alberta, Canada}
\affil[4]{Department of Physics and Astronomy, University of Alabama, Tuscaloosa, USA}
\affil[5]{IFIC, Universitat de Valencia-CSIC, Valencia, Spain}

\date{} 

\begin{document}
\maketitle

\begin{abstract}
The calibration of the CR39\textsuperscript{\textregistered} and Makrofol\textsuperscript{\textregistered} Nuclear Track Detectors of the MoEDAL experiment at the CERN-LHC was performed by exposing stacks of detector foils to heavy ion beams with energies ranging from 340 MeV/nucleon to 150 GeV/nucleon. After chemical etching, the base areas and lengths of etch-pit cones were measured using automatic and manual optical microscopes. The response of the detectors, as measured by the ratio of the track-etching rate over the bulk-etching rate, was determined over a range extending from their threshold at Z/$\beta\sim7$ and $\sim50$ for CR39 and Makrofol, respectively, up to Z/$\beta\sim92$.
\end{abstract}

\noindent\textbf{Keywords:} Particle tracking detectors, Detector alignment and calibration methods (particle-beams), Particle identification methods.

\flushbottom

\maketitle
\flushbottom

\section{Introduction}
\label{sec:intro}
Solid State Nuclear Track Detectors (SSNTDs) are widely used in several scientific applications, such as nuclear fragmentation studies~\cite{fragmentation}, cosmic ray composition and exotic particles searches in the cosmic radiation and at particle accelerators~\cite{CAKE,SLIM,BI,isr1,isr2,isr3}. They are often the detectors of choice when it comes to the search for highly ionizing particles in a background dominated by minimum ionizing particles. 

\begin{figure}[htb!]
\includegraphics[width =1.\hsize,clip]{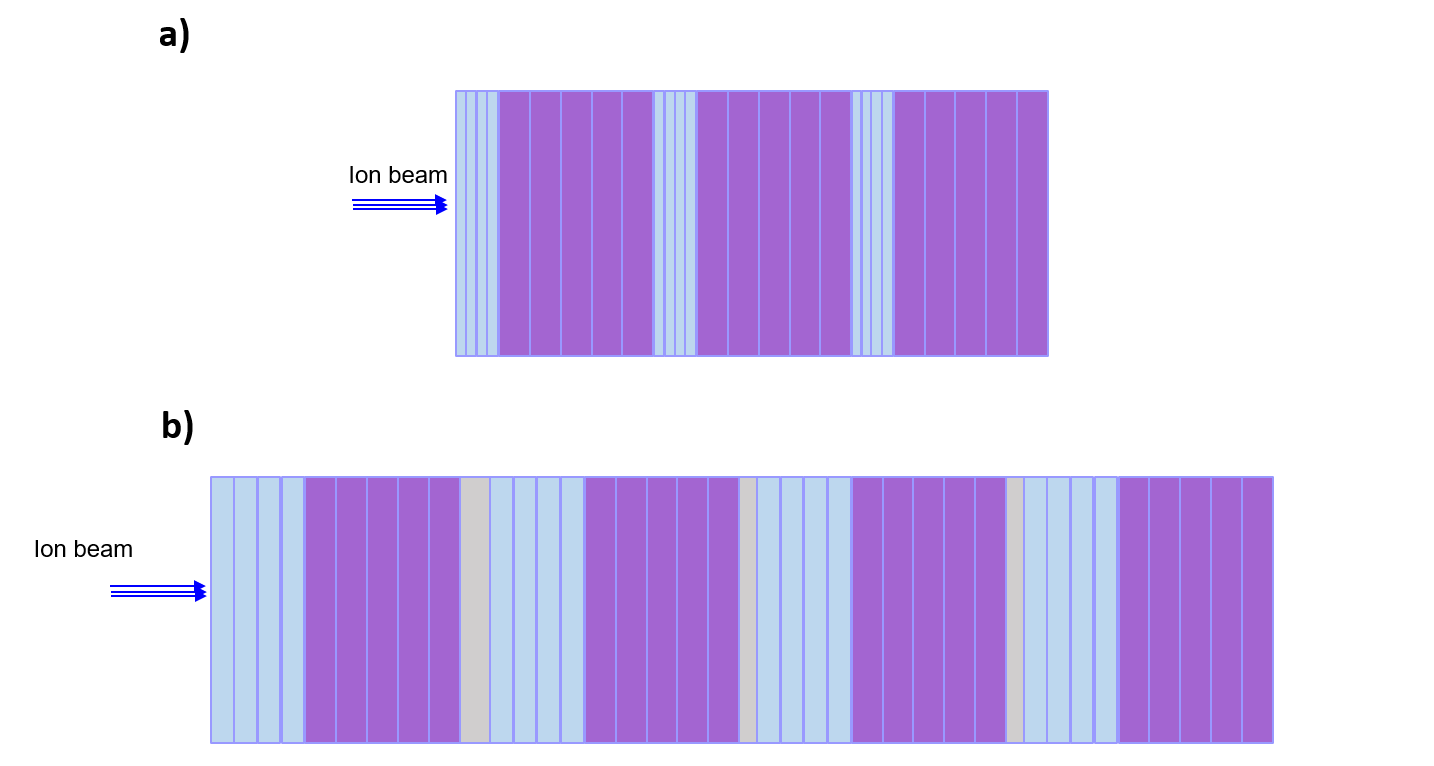}
\caption{\label{fig:stacks}Scheme of calibration stacks: a) stack composed of several CR39 foils; b) stacks with one or two 9.5 mm thick aluminum slabs interleaved to groups of SSNTDs foils. In blue are CR39 foils 0.5 or 1 mm thick, in purple are CR39 foils 1.5 mm thick  and in grey are the aluminum slabs. }
\end{figure}  
A charged particle crossing a foil of plastic SSNTD  breaks the polymeric bonds in a narrow (O(10 nm)) cylindrical region (``latent track") along its trajectory. By a suitable chemical etching process, which acts preferentially along the  latent track, conical ``etch-pits" are formed. Highly ionizing particles will exhibit a distinct signature characterized by continuous tracks with significant and consistent dimensions through multiple layers of SSNTDs.

 \par Calibrations of CR39 and Makrofol 
 have been performed in the past~\cite{giaco, calib07}. 
Since the response of SSNTDs depends very sensitively  on applied etching conditions, a re-calibration becomes necessary whenever the detectors are deployed for a new experiment. 
 
 \par In this paper we report the results of the calibration of the CR39 manufactured by TASL Ltd, and, Makrofol produced by Bayer Films LLC, which are both used in the SSNTD sub-detector of the MoEDAL experiment at the Large Hadron Collider (LHC) located at the European Centre for Nuclear Research (CERN). MoEDAL is  designed to search for Highly Ionizing Particle (HIPs) such as magnetic monopoles and highly electrically charged particles \cite{moedal1,moedal2}.  
In MoEDAL stacks of CR39 and  Makrofol foils are deployed over a surface  of about 10 m$^2$ around the Interaction Point (IP8) shared with the LHCb experiment.  

\begin{table}
  \caption{Ions used for the calibration of CR39 at NSRL. Z$/\beta$ and REL are the charge to velocity ratio, and the restricted energy loss of the beam ions, respectively.}
  \vspace{0.5cm}
  \label{tab:ionbeams}
\centering
  \begin{tabular}{|l|c|c|c|}
    \hline
     Ions  & Kinetic energy   &   Z$/\beta$ & REL \\
     & (MeV/nucleon) &     & (MeVg$^{-1}$cm$^2$)\\
    \hline
      $^{12}$C$^{6+}$ & $1000$ & 6.9 & 42.6\\
      $^{16}$O$^{8+}$ & $1000$ & 9.2 & 75.8\\
      $^{28}$Si$^{14+}$ & $1000$ & 16.2 & 232.2\\
      $^{28}$Si$^{14+}$ & $715$ & 17.2 & 256.3\\
      $^{28}$Si$^{14+}$ & $555$ & 18.3 & 281.9\\
      $^{28}$Si$^{14+}$ & $450$ & 19.3 & 309.2\\
      $^{56}$Fe$^{26+}$ & $1000$ & 30.0 & 800.8\\
      $^{56}$Fe$^{26+}$ & $340$ & 39.1 & 1232.2\\
      $^{84}$Kr$^{36+}$ & $383$ & 52.1 & 2215.6\\
      $^{129}$Xe$^{54+}$ & $350$ & 80.4 & 5209.4\\
    \hline
  \end{tabular}
\end{table}

\section{Experimental Method}
\label{sec:experiment}
The calibration  process adopted for the MoEDAL SSNTDs involves the exposure of the detectors   to  heavy ion beams at particle accelerators. Irradiated foils undergo chemical etching in aqueous solutions of KOH or NaOH, at concentrations typically  of 4N - 6N, and temperatures of  50 - 80~$^{\circ}$C, resulting in the formation of conical etch-pits. The shape and size of etch-pits depend on the particle's restricted energy loss (REL), its angle of incidence and the etching condition. 

\subsection{Exposure and Etching  of CR39}
Stacks of SSNTD layers composed of  11.5 cm $\times$ 11.5 cm foils,  0.5, 1.0 and 1.5 mm thick, were exposed in 2021 at the NASA Space Radiation Laboratory (NSRL) to $^{12}$C$^{6+}$, $^{16}$O$^{8+}$, $^{28}$Si$^{14+}$, $^{56}$Fe$^{26+}$, $^{84}$Kr$^{36+}$, and $^{129}$Xe$^{54+}$ ion beams with energies ranging from 340 MeV/nucleon to 1 GeV/nucleon. The initial values of ions' kinetic energy, charge on $\beta$ ratio (Z/$\beta$) and restricted energy loss (REL) are listed in Table~\ref{tab:ionbeams}. Exposures were performed with ions impinging normally on the detector surface. 
\par The exposed stacks comprised layers of CR39 and other  SSNTDs such as Makrofol and PET (Figure~\ref{fig:stacks}a). In order to slow down incident ions,  slabs of aluminum, $9.5$~mm and  $19$~mm thick (Figure~\ref{fig:stacks}b) were introduced in some stacks.  Passing through the plastic foils and the aluminium slabs, ion's kinetic energy and charge Z decrease, while the value of REL and ratio Z/$\beta$ increase. Since along their path ions can undergo charge-changing fragmentation, with  a single exposure it is possible to probe the response of CR39  to a wide range of nuclear charge and energy losses.

\begin{figure}[htb!]
\vspace{1.0cm}
\includegraphics[width = 0.47\hsize,clip]{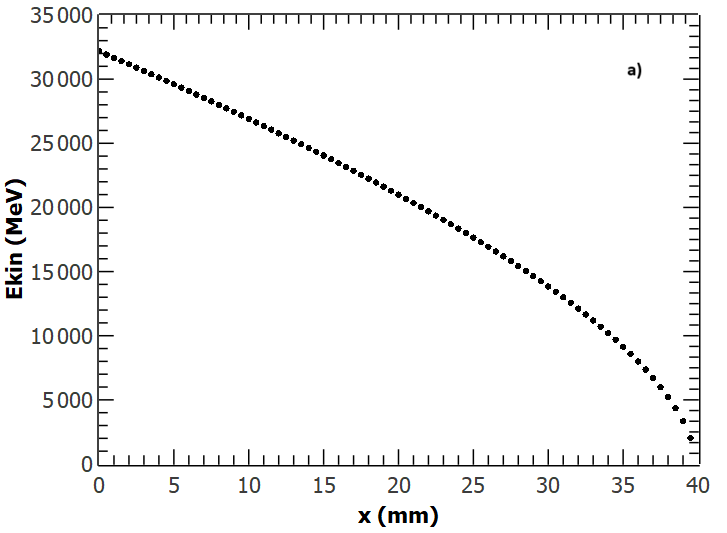}
\includegraphics[width = 0.47\hsize,clip]{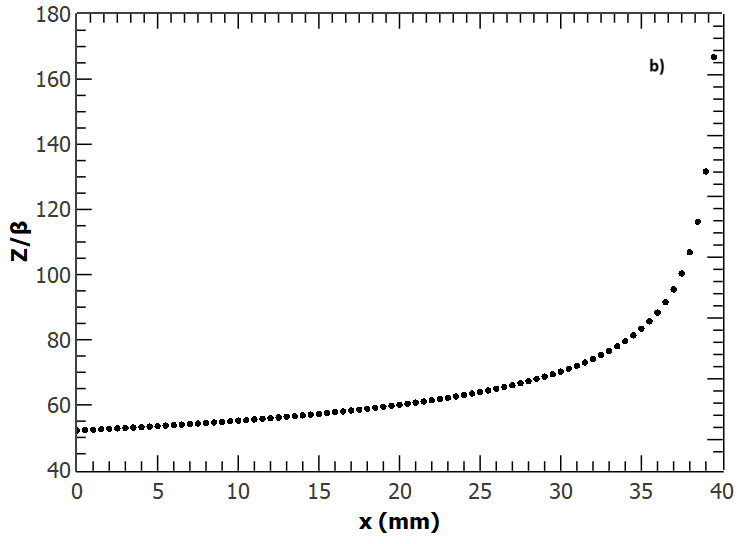}
\caption{\label{fig:kr} Kinetic energy (a) and (b) Z/$\beta$ ratio of Kr ion across a 40 mm thick CR39 foils}
\end{figure}

\begin{figure}[htb]
\includegraphics[width = 0.47\hsize,clip]{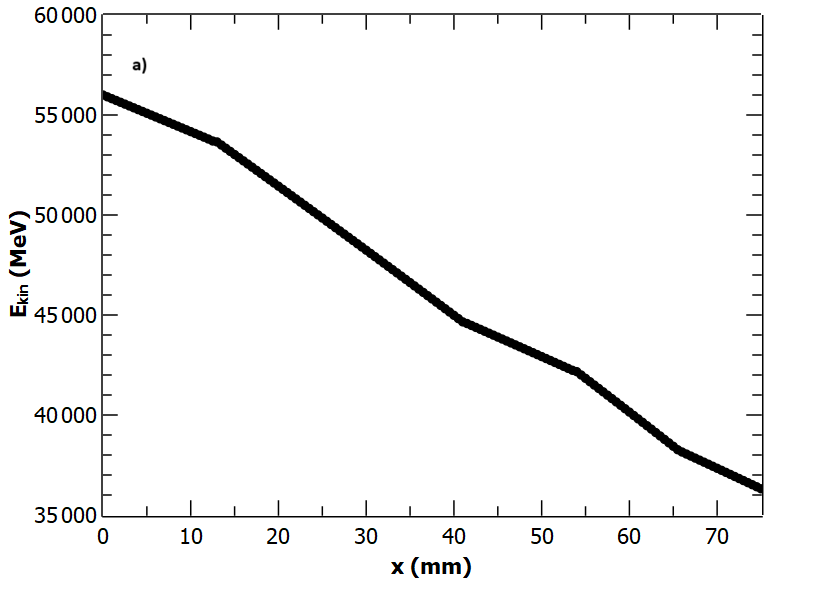}
\includegraphics[width = 0.47\hsize,clip]{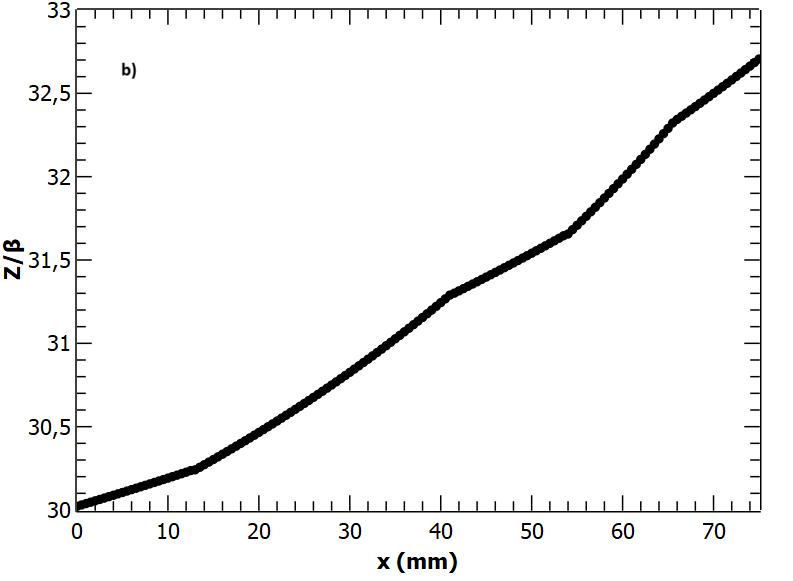}
\caption{\label{fig:fe} Kinetic energy (a) and (b) Z/$\beta$ of Fe ion across 70 mm thick stack composed of CR39 foils and aluminum slabs. The change in slope corresponds to  transitions from a group of SSNTD foils to an Aluminum slab.}
\end{figure}

Figure~\ref{fig:kr}a shows the variation of $^{84}$Kr$^{36+}$ ions' kinetic energy along a stack of CR39 foils. Figure~\ref{fig:kr}b shows, for the same ions, the Z/$\beta$ versus depth in the stack. Figures~\ref{fig:fe}a,b  show the kinetic energy and Z/$\beta$, respectively, of $^{56}$Fe$^{26+}$ ions along a stack composed of CR39 foils and aluminum slabs. 

\begin{figure}[ht]
\centering
\includegraphics[width =0.8\hsize,clip]{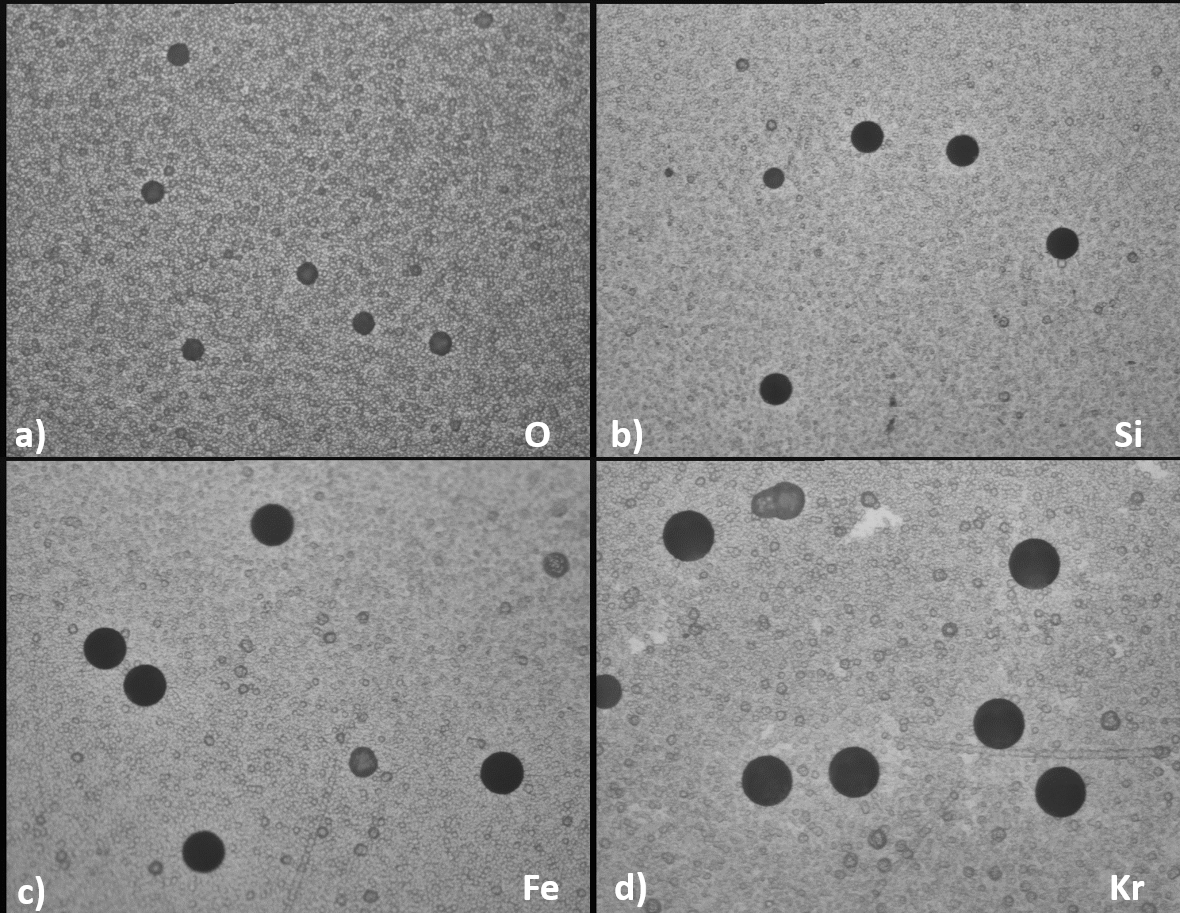}
\caption{\label{fig:track} Ion tracks in CR39 after 15 hours etching in 6 N NaOH solution at 70 $^{\circ}$C as seen under an optical microscope with $20\times$ objective, coupled to a CCD camera: from top left clockwise: Oxygen ions, Z/$\beta$ = 9; Silicon ions, Z/$\beta$ = 24; Krypton ions, Z/$\beta$ = 52 and Iron ions, Z/$\beta$ = 39.}
\end{figure}

After exposure, the CR39 foils were etched in 6N NaOH solution at (70.0 $\pm~$0.1)$^{\circ}$C for 15~h. 
 For ions impinging normally to the surface, as in the present case, surface openings of etch-pits observed under an optical microscope  manifest as dark circles, referred to also as ``tracks''. Figure~\ref{fig:track} shows  tracks of $^{16}$O, $^{28}$Si, $^{56}$Fe and $^{84}$Kr ions in  CR39.

The CR39 bulk etching rate, $v_B$, was determined for each detector foil, by measuring its thickness, before and after etching, with an accuracy of 1$\mu$m on a grid of 20 points.
The average $v_B$ is (1.27$~\pm~$0.05$)~\mu$m/h. The $v_B$ of the foils ranges from (1.1$~\pm~$0.05$)~\mu$m/h up to (1.5$~\pm~$0.05$)~\mu$m/h.

\subsection {Exposure and Etching of Makrofol}

The process is very similar to the one applied to the CR39. The main difference concerns the etchant. It has been observed that for Makrofol, the inclusion of Ethyl Alcohol in the etching solution has a polishing effect on the Makrofol surface, which results in a smoother surface and sharper etch-pit edges. 
 Commercial  Ethyl Alcohol contains 1 \% of additives, such as methyl ethyl ketone (MEK), isopropyl alcohol (IPA), and denatonium benzoate introduced during the denaturing process. The types and amount of additives in Ethyl Alcohol were changed  according to EU regulations issued in 2013 and in 2017. Experimental investigations revealed that Ethyl Alcohol prepared according to the 2013 EU regulations yielded the most favorable etching results. Since then, Makrofol detectors were etched using this specific type of Ethyl Alcohol. An agreement has been established with the manufacturer to ensure a consistent supply of this particular variety.

 Several stacks of Makrofol foils measuring 11.5 cm $\times$ 11.5 cm and with a thickness of 500~$\mu$m were exposed to 156 GeV/nucleon Pb$^{82+}$  and 13 GeV/nucleon Xe$^{54+}$ nuclei at the CERN-SPS in 2017 and 2018. Upstream foils in the stacks served both as a detector and fragmentation target; downstream foils recorded both Pb and Xe ions along with their fragments.
 
\begin{figure}[htb]
\centering
\includegraphics[width =0.9\hsize,clip]{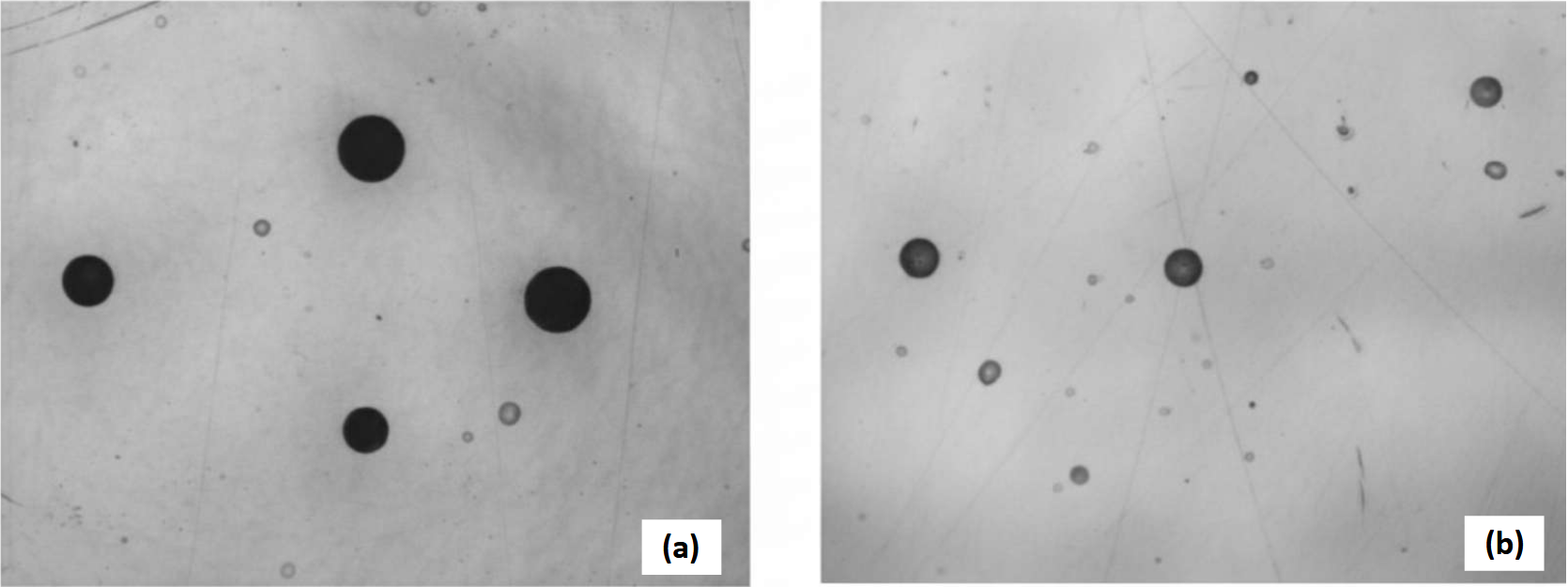}
\caption{\label{fig:maktracks} Tracks of (a) Pb and (b) Xe ions, along with their nuclear fragments, observed in Makrofol after 15
hours of etching in a 5.5N KOH solution with 20~\% ethyl alcohol at 45$^{\circ}$C. The dimensions of the image frames are 0.80 mm by 0.64 mm.}
\end{figure} 

 After exposure Makrofol foils were etched in 5.5N KOH + 20~\% Ethyl Alcohol at (45.0 $ \pm~ $0.1)$^{\circ}$C for 15~h. The bulk etching rate, determined by measuring the mean thickness of the foils before and after etching was $v_B$ = (3.01 $\pm$ 0.05) $\mu$m/h. 
The application of the selected etching condition resulted in well-defined conical etch-pits with sharp edges against a clear surface as shown in figure~\ref{fig:maktracks}. 

\section{Etch-pit measurements}
\label{sec:results}
\begin{figure}[ht]
\includegraphics[width = 1.0\hsize,clip]{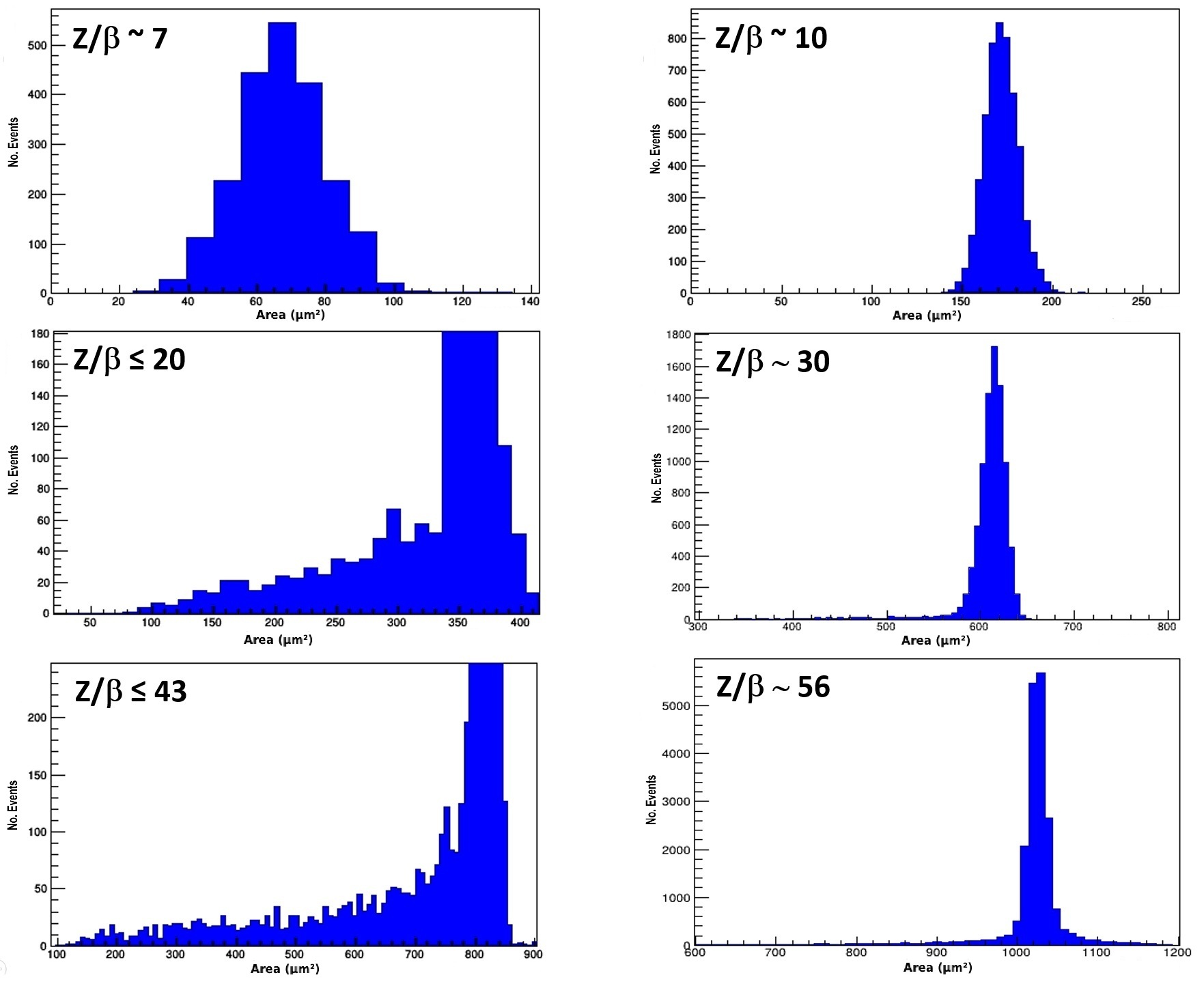}
\caption{\label{fig:areacr} Distributions of etch-pits base area measured in CR39 foils, exposed to Carbon ions (upper left), Oxygen ions (upper right), Silicon ions (center left), Iron ions (center right and lower left) and Krypton ions (lower right). Distributions are obtained from foils exposed to different ions, and for different positions in the stacks, first layers for Carbon, Oxygen an Iron (center right) ions and innermost layers for Silicon , Iron (lower left) and Krypton ions.}
\end{figure}

After etching, the base areas of conical etch-pits in CR39 and Makrofol foils were measured using an automated microscope featuring a stage with automated movements in the X and Y directions and automatic focusing on each frame. It was coupled to a computer equipped with image analysis software. The distributions of track areas in CR39 foils for  different ion species are shown in figure~\ref{fig:areacr}. 

\begin{figure}[htb]
\includegraphics[width = 1.0\hsize,clip]{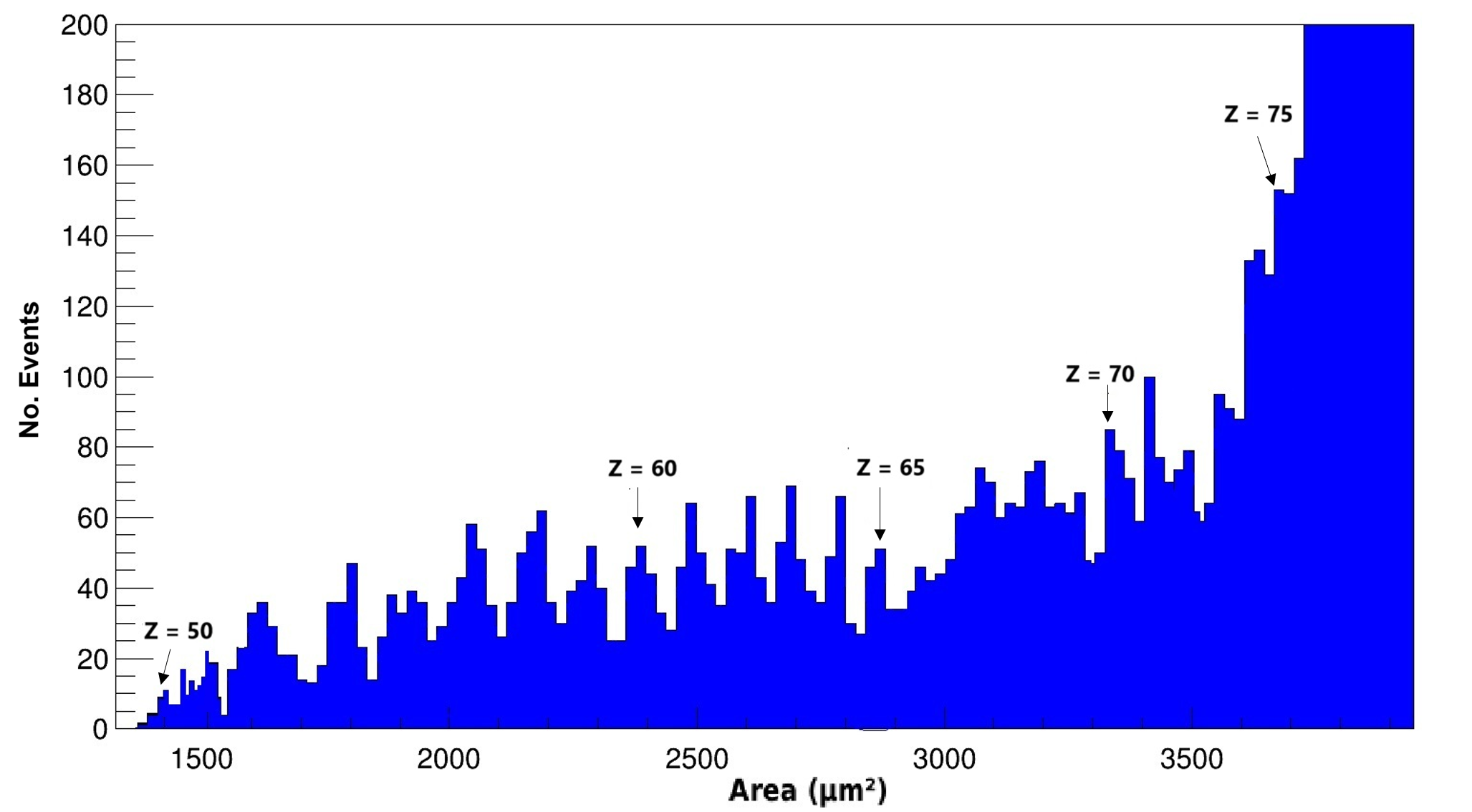}
\caption{\label{area} Etch-pit base area distribution of lead ions and charged nuclear fragments in Makrofol.}
\end{figure}

Figure~\ref{area} illustrates the area distribution of lead ion tracks and their fragments in Makrofol. The distributions in figure~\ref{fig:areacr} and in figure~\ref{area} are obtained by averaging the measurements of etch-pit surface areas on both the front and back surfaces of two or three successive foils in a stack after correctly aligning the reference frames of all surfaces to a unique one. This procedure enhances charge resolution.

\begin{figure}
\centering
\includegraphics[width = 0.9\hsize,clip]{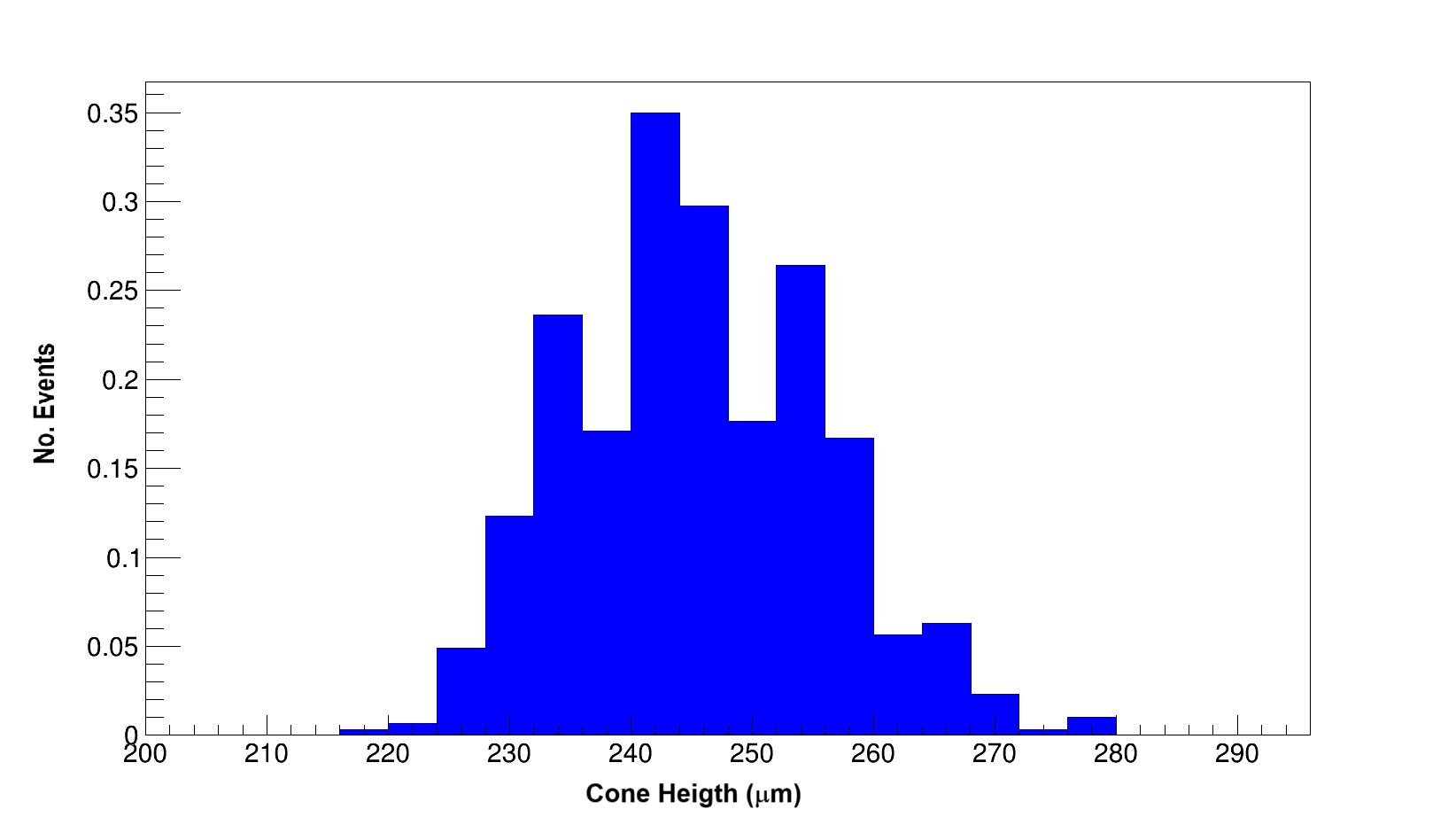}
\caption{\label{fig:depth} Normalized distribution of etch-pit lengths of $^{129}$Xe ions and their fragments in CR39. The distribution is obtained by combining data from two successive foils with different measurement statistics, resulting in a total of slightly less than 500 measurements.}
\end{figure}

\begin{figure}[htb]
\centering
\includegraphics[width = 0.9\hsize,clip]{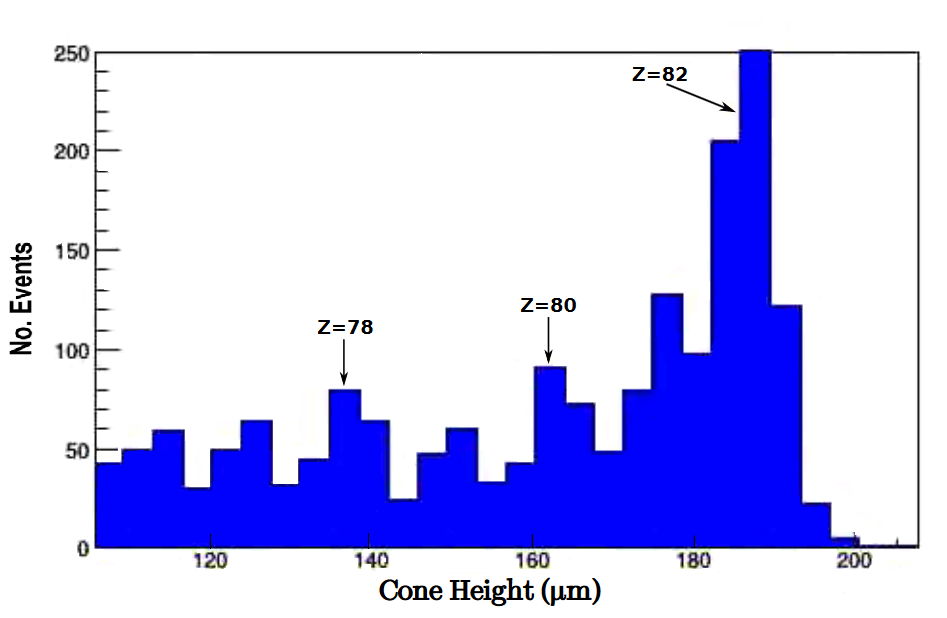}
\caption{\label{tracklen}The etch-cone length distribution for Pb and its fragments on Makrofol, showing well defined peaks for large ion charges.}
\end{figure}

Both in CR39 and in Makrofol the charge resolution decreases for Z/$\beta$$\geq$75. For  larger Z/$\beta$ values, charge  resolution can be recovered by  measuring the etch-cone length. For the calibration of Makrofol, etch-pits with the largest surface areas were selected, and their lengths were manually measured using a Leica DMRME microscope with a resolution of $1\mu$m. Figure~\ref{fig:depth} and figure~\ref{tracklen} show the distributions of etch-pit lengths in CR39 and in Makrofol, respectively.

 \section{The reduced etch-ratio versus REL} 
 \par  The  ratio  $p$ = $v_T$/$v_B$, where $v_T$ is the etching rate along the latent track, is used to evaluate the detector response (referred to as ``etch-rate'' in the following).  The etch-rate was determined using eq. \eqref{eq:p_area} if the surface track area $A$ was available, and eq. \eqref{eq:p_length} if the etch-pit length, $L$, had been measured~\cite{giaco}. 
In both equations $t$ is the etching duration. In eq. \eqref{eq:p_length} $L_e$ is the cone length obtained by taking into account the CR39 or Makrofol refractive index ($\sim 1.498$ and $\sim 1.58$, respectively). 

\begin{equation}
\label{eq:p_area}
p = \frac{1+\left(\frac{A/\pi}{(v_B t)^2}\right)}{1-\left(\frac{A/\pi}{(v_B t)^2}\right)} 
\end{equation}

\begin{equation}
\label{eq:p_length}
p = 1 +\frac{L_{e}}{v_{B} t} 
\end{equation}
 
\begin{figure} [htb]
\includegraphics[width = 0.9\hsize,clip]{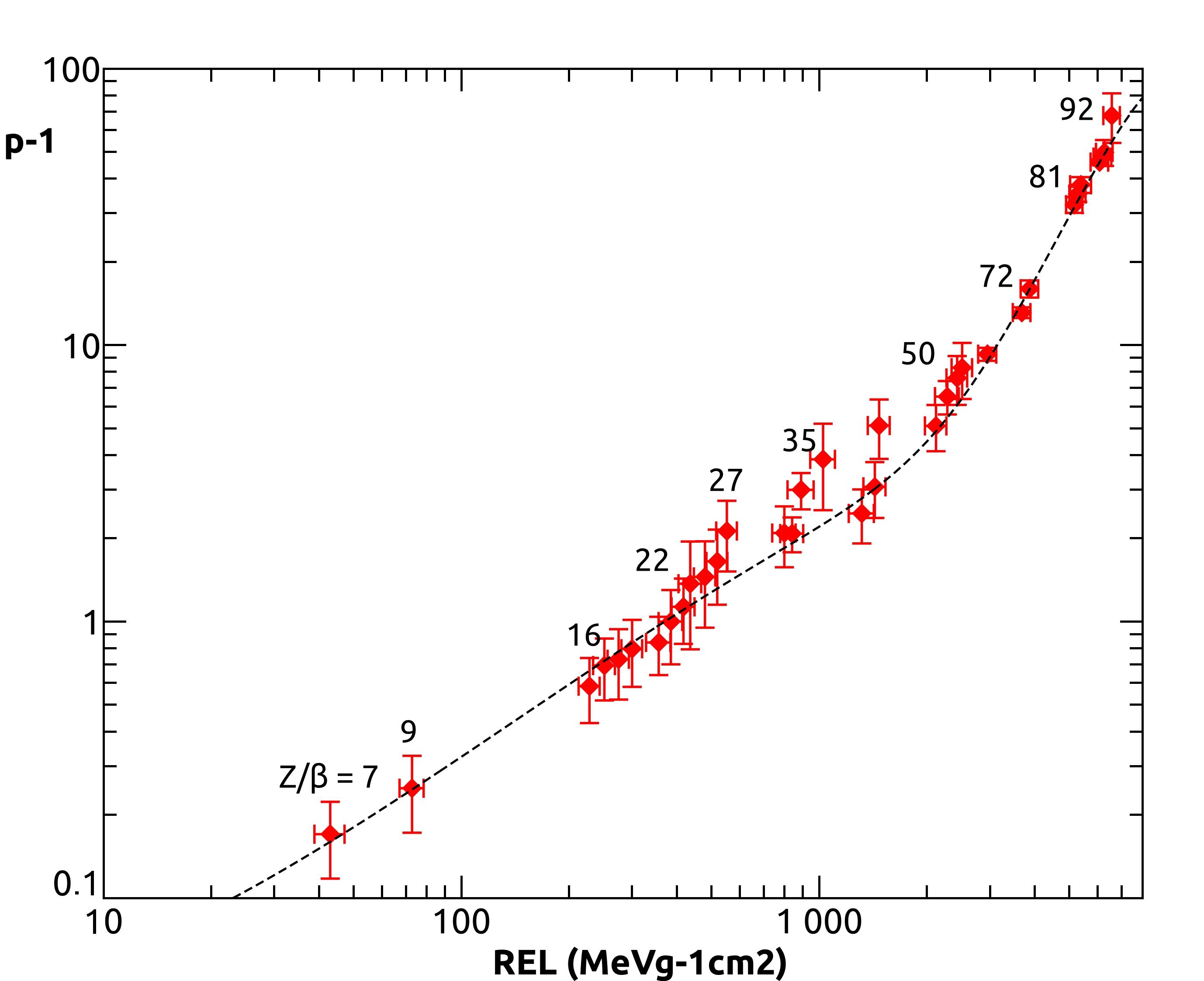}
\caption{\label{fig:calibration}  Reduced etch-rate of CR39 versus REL. The dashed curve is a 4-degree polynomial fit  to the data (see text).}
\end{figure}

A Monte Carlo simulation of the ions' propagation through the stacks was implemented. The energy loss was computed using the SRIM~\cite{SRIM} package. The ion's energy  was evaluated after every detector layer or aluminum absorber with $\sim$ 10 \% accuracy. The  reduced etch rate $p-1$ versus REL for CR39 is plotted in figure~\ref{fig:calibration}. The detector's threshold, defined as the value of REL corresponding to p = 1, is at $\sim$ 40 MeVg$^{-1}$cm$^{-2}$, which corresponds to the restricted energy loss  of relativistic Z/$\beta$ = 7 Nitrogen  ion. Calibration data extend up to an equivalent relativistic charge of $\sim$92.
The dashed curve in figure~\ref{fig:calibration} is a 4-degree polynomial fit:
\begin{equation}
\label{eq:fit}
  p-1 = a_0 + a_1\times REL + a_2 \times REL^2 + a_3 \times REL^3 + a_4 \times REL^4,
\end{equation}

\noindent with parameters: $a_0 = 0.0297 \pm  0.059$, $a_1 = (3.08 \pm 0.44)\times 10^{-3}$, $a_2 = (-1.47 \pm 0.52)\times 10^{-6}$, $a_3 = (5.96 \pm 1.77) \times 10^{-10}$ and $a_4 = (-3.84 \pm 1.71)\times 10^{-14}$,  and $\chi^2/dof = 1.28$.
 
\begin{figure}[htb]
\includegraphics[width = 0.9\hsize,clip]{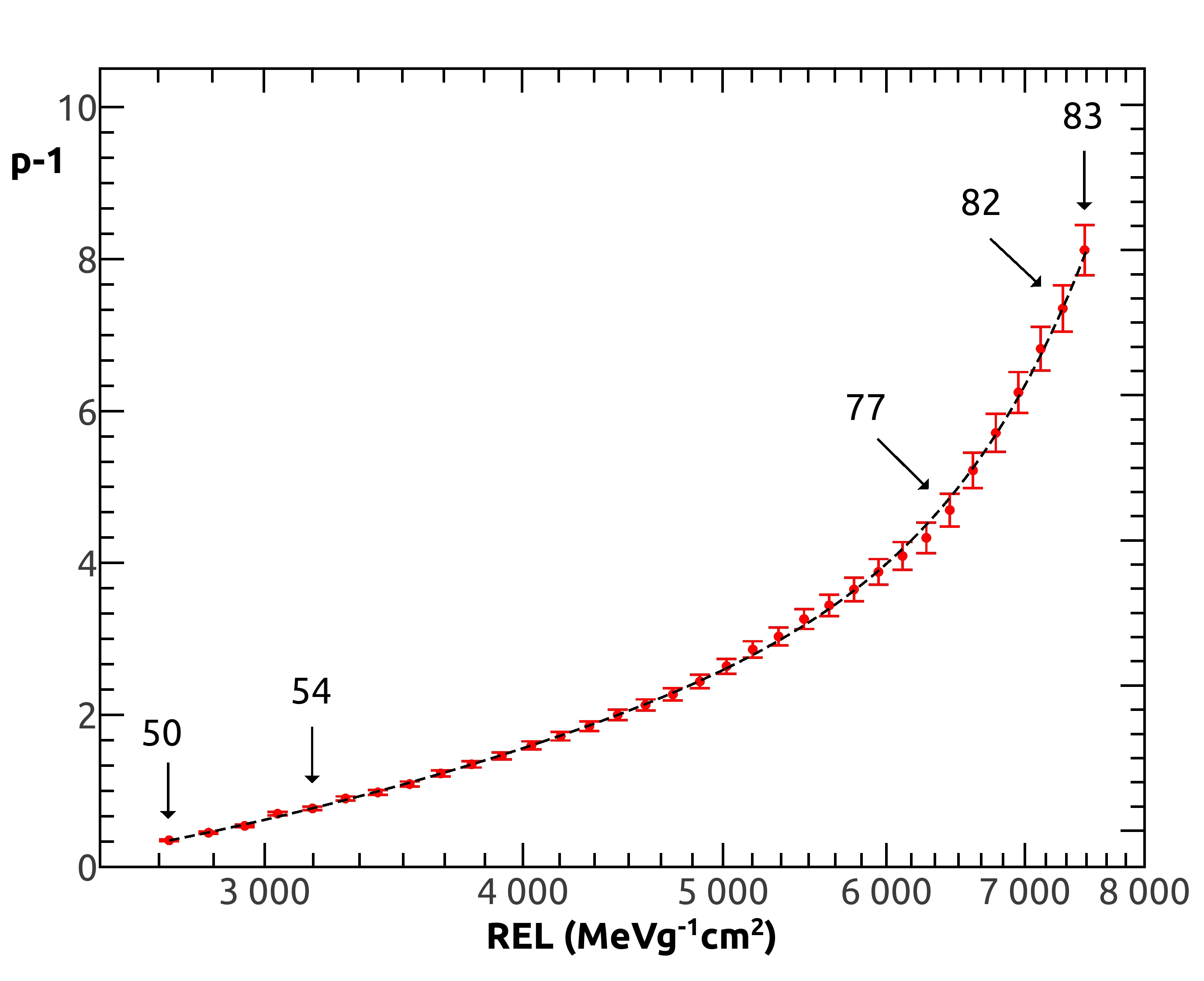}
\caption{\label{calibration}  Reduced etch rate p-1 versus REL for  Makrofol SSNTD exposed to relativistic lead and xenon ion beams. The dashed curve represents a polynomial fit.}
\end{figure}

The reduced etch-rate $p - 1$ versus REL for the Makrofol SSNTD is shown in figure~\ref{calibration}. The detection threshold is at $\sim$ 2700 MeVg$^{-1}$cm$^{-2}$, corresponding to a relativistic nuclear fragment with charge Z/$\beta$ $\sim$~50. 
\noindent The dashed curve is a 4-degree polynomial fit, as in eq.~\ref{eq:fit}, with parameters: $a_0 = (-1.37 \pm  1.03) \times 10^{-6}$, $a_1 = (-1.42 \pm 1.00)\times 10^{-3}$, $a_2 = (9.61 \pm 3.52)\times 10^{-7}$, $a_3 = (-1.76 \pm 0.53) \times 10^{-10}$ and $a_4 = (1.24 \pm 0.29)\times 10^{-14}$,  and $\chi^2/dof = 0.3$.
 
\section{Conclusions}
\label{sec:concl}
The response of CR39 and Makrofol solid state nuclear track detectors used in the MoEDAL experiment was calibrated using several ions at different energies covering a large Z/$\beta$ range.  The CR39 can detect particles with a restricted energy loss as low  as 40 MeV g$^{-1}$cm$^2$, or equivalently with  Z/$\beta$ = 7,  up to  REL $\sim 7000$ MeV g$^{-1}$cm$^2$, or equivalently Z/$\beta$ = 92.
Makrofol on the other hand is shown to have a higher detection threshold and can only detect particles with a restricted energy loss of at least 2700 MeV g$^{-1}$cm$^2$ or equivalently Z/$\beta$ = 50.

In conclusion, both CR39 and Makrofol are shown to be excellent choices as detectors for a wide range of applications where the detection of heavy ions against a low-Z background is important, such as in Nuclear Physics fragmentation, Cosmic Ray composition, etc. They are also important in the search for heavily ionizing rare events, for which a clear signature is expected over all possible backgrounds. CR39 and Makrofol have already made a significant contribution to setting limits on the production of predicted charged exotic particles and can  potentially contribute to a discovery.

\section*{Acknowledgments} 
 The authors sincerely thank the staff at CERN-SPS and  NSRL for their help during the irradiation of the SSNTD stacks. The work of A. Maulik was supported by the University of Alberta, Canada and that of A. Upreti by the NSF grant 2309505 of the MoEDAL group at the University of Alabama, USA.

\end{document}